\documentclass[iop]{emulateapj}

\usepackage{ifthen}
\newboolean{aastex_on}
\setboolean{aastex_on}{false}
\newboolean{electronic_on}
\setboolean{electronic_on}{true}

\usepackage{amsmath}
\usepackage{amssymb}
\usepackage{graphicx}
\usepackage{color}
\usepackage{array}
\shorttitle{Detection and Imaging of the Crab with NCT}
\shortauthors{Bandstra et al.}

\begin{document}

\title{Detection and Imaging of the Crab Nebula with the Nuclear Compton Telescope}


\author{M.~S.~Bandstra\altaffilmark{1}, E.~C.~Bellm\altaffilmark{1}, S.~E.~Boggs\altaffilmark{1}, D.~Perez-Becker\altaffilmark{1}, A.~Zoglauer\altaffilmark{1}, H.-K.~Chang\altaffilmark{2}, J.-L.~Chiu\altaffilmark{2}, J.-S.~Liang\altaffilmark{2}, Y.-H.~Chang\altaffilmark{3}, Z.-K.~Liu\altaffilmark{3}, W.-C.~Hung\altaffilmark{3}, M.-H.~A.~Huang\altaffilmark{4}, S.~J.~Chiang\altaffilmark{4}, R.-S.~Run\altaffilmark{4}, C.-H.~Lin\altaffilmark{5}, M.~Amman\altaffilmark{6}, P.~N.~Luke\altaffilmark{6}, P.~Jean\altaffilmark{7}, P.~von~Ballmoos\altaffilmark{7}, C.~B.~Wunderer\altaffilmark{8}}

\altaffiltext{1}{University of California Space Sciences Laboratory, Berkeley, CA 94720}
\altaffiltext{2}{National Tsing Hua University, Hsinchu, Taiwan, 30013}
\altaffiltext{3}{National Central University, Taoyuan, Taiwan, 32001}
\altaffiltext{4}{National United University, Miaoli, Taiwan, 36003}
\altaffiltext{5}{Institute of Physics, Academia Sinica, Taipei, Taiwan, 11529}
\altaffiltext{6}{Lawrence Berkeley National Laboratory, Berkeley, CA 94720}
\altaffiltext{7}{Centre d'Etude Spatiale des Rayonnements (CESR), 31028 Toulouse, France, Cedex 4}
\altaffiltext{8}{Deutsches Elektronen-Synchrotron (DESY), Notkestr. 85, 22607 Hamburg, Germany}

\email{Electronic address: bandstra@ssl.berkeley.edu}

\begin{abstract}
The Nuclear Compton Telescope (NCT) is a balloon-borne Compton telescope designed for the study of astrophysical sources in the soft gamma-ray regime (200~keV--20~MeV)\@. NCT's ten high-purity germanium crossed-strip detectors measure the deposited energies and three-dimensional positions of gamma-ray interactions in the sensitive volume, and this information is used to restrict the initial photon to a circle on the sky using the Compton scatter technique\@. Thus NCT is able to perform spectroscopy, imaging, and polarization analysis on soft gamma-ray sources\@.  NCT is one of the next generation of Compton telescopes --- so-called compact Compton telescopes (CCTs) --- which can achieve effective areas comparable to COMPTEL's with an instrument that is a fraction of the size\@. The Crab Nebula was the primary target for the second flight of the NCT instrument, which occurred on 17--18 May 2009 in Fort Sumner, New Mexico\@.  Analysis of 29.3~ks of data from the flight reveals an image of the Crab at a significance of $4\sigma$.  This is the first reported detection of an astrophysical source by a CCT\@.
\end{abstract}

\keywords{gamma rays: general --- ISM: individual (Crab Nebula) --- Techniques: imaging spectroscopy --- Methods: data analysis --- Telescopes --- Balloons}

\section{Introduction}

Soft gamma rays (200~keV--20~MeV) are important probes of the most violent and extreme processes in the cosmos\@. Gamma rays are produced by non-thermal processes in such disparate objects as neutron stars, X-ray binaries, and Active Galactic Nuclei (AGN), and they also result from the decays of many radioactive nuclei that are released by supernova explosions, such as $^{26}$Al, $^{60}$Fe, and $^{44}$Ti\@. The penetrating nature of gamma rays allows the astrophysicist to probe deep within these often obscured systems and make unique and complementary observations of their gravitational fields, magnetic fields, and nuclear reactions\@.

Over the last four decades, various types of telescopes have been developed to detect and image soft gamma-rays\@. One prominent technology is the Compton telescope, which exploits the Compton effect to perform direct imaging of gamma-ray photons\@.  The Compton effect occurs when a gamma-ray photon incoherently scatters with an electron and is deflected by an angle $\varphi$\@.  The Compton scatter formula relates the energy of the scattered gamma-ray photon $E_{\gamma_{1}}$ and the incident photon energy $E_{\gamma_{0}}$ to the scatter angle $\varphi$:
\begin{equation}
	\cos\varphi = 1 - m_{e}c^{2} \left(\frac{1}{E_{\gamma_{1}}}-\frac{1}{E_{\gamma_{0}}}\right) 
\label{eq:compton_scatter_formula}
\end{equation}
where $m_{e}c^{2}=511$~keV is the rest mass energy of the electron \citep{Compton:1923}\@. The basic principle of Compton imaging is to measure the energies deposited by at least one Compton scatter and employ Equation~\ref{eq:compton_scatter_formula} to reconstruct the direction of the incoming photon\@.  If both the direction of the scattered photon and $\cos\varphi$ can be measured, then the photon's direction can be constrained to a circle on the sky (the ``Compton circle'')\@.  The Compton effect is also intrinsically sensitive to the polarization of the incident photon through the Klein-Nishina cross section \citep{Klein:1929:p853}, enabling a Compton telescope to potentially measure the polarization of a gamma-ray source\@.

The first satellite-borne Compton telescope was COMPTEL on board the Compton Gamma Ray Observatory~\citep{Schonfelder:1993:p657}, which consisted of two layers of scintillator detectors spaced by 1.5~meters\@. Photons were required to scatter in the top layer and be fully absorbed in the bottom layer\@. COMPTEL succeeded in making measurements and images of the gamma-ray regime to unprecedented sensitivity, including the first detection of $^{44}$Ti from a supernova remnant~\citep{Iyudin:1994:pL1} and a map of the galactic $^{26}$Al distribution~\citep{Oberlack:1996:p311}\@.

To build on the successes of COMPTEL, the next generation Compton telescopes utilize a compact design, which requires detectors with both fine position and fine energy resolution to track individual gamma-ray interactions within the detector volume\@. Compact Compton Telescopes (CCTs) can achieve effective areas comparable to COMPTEL's with an instrument that is a fraction of the size~\citep{Boggs:2001:p1126}\@. The first CCT was LXeGRIT, which uses a single liquid xenon time projection chamber (TPC) to measure interaction positions with high spatial resolution ($\approx$3~mm) and energy resolution ($\approx$10\% at 1~MeV)~\citep{Aprile:1993:p279, Aprile:2000:p333}\@. Various CCTs are currently being developed that use other detector materials (e.g.,~\cite{Kurfess:2007:p367, ONeill:2003:p251, Bloser:2002:p611, Takeda:2007:p19, Ueno:2008:p96, Hunter:2008:p1389, Kroeger:1995:p236, Kurfess:2003:p256})\@.

In this paper we will briefly describe the NCT instrument (Section~\ref{sec:instrument}), then describe the event selections and present the first image of the Crab (Sections~\ref{sec:data_selections} and \ref{sec:image})\@.  We discuss two analysis tools: Monte Carlo simulations of our observation of the Crab and the Angular Resolution Measure (ARM) for a Compton telescope (Sections~\ref{sec:sim} and \ref{sec:arm_def})\@. Using these tools, we determine the significance of the Crab detection using the total counts, an analysis of the spectrum, and an analysis of the ARM histogram (Sections~\ref{sec:significance}--\ref{sec:analysis_arm})\@.  Polarization analysis of the Crab was not performed due to the lack of sufficient counts\@.

\section{The NCT instrument}

\label{sec:instrument}

The NCT instrument~\citep{Boggs:2001:p877,Boggs:2003:p1221,Boggs:2004:p251,Chang:2007:p1281} is a balloon-borne CCT that utilizes high-purity crossed-strip germanium detectors for Compton imaging\@. To date, NCT has flown on two conventional balloon flights\@. The first was a prototype flight in 2005 that succeeded in measuring the soft gamma-ray atmospheric background and galactic anticenter region~\citep{Coburn:2005:p13,Bowen:2007:p436,Bowen:2009:Thesis}\@. The full 10-GeD version of NCT was flown on its second flight in May 2009~\citep{Bandstra:2009:p2131}, and the first results are presented here\@.

The heart of NCT is its ten crossed-strip high-purity germanium detectors (GeDs)~\citep{Amman:2000:p155,Amman:2007:p886}\@. Each GeD has the ability to record in three dimensions the location of each individual photon interaction\@. A single GeD measures 8~cm $\times$ 8~cm $\times$ 1.5~cm and has 37 strips on each side, with a strip pitch of 2~mm\@. NCT is designed such that when a gamma ray undergoes at least one Compton scatter and is fully absorbed inside the detectors, the individual scatter positions and energy deposits can be measured\@. Compton Kinematic Discrimination algorithms~\citep{Aprile:1993:p216, Boggs:2000:p311, Oberlack:2000:p168, Kroeger:2002:p1887} are used to find the most likely scattering order, and then the total gamma-ray energy can be reconstructed and the initial direction can be traced to a ring on the sky\@. Thus NCT works as both a spectrometer and an imager, and is intrinsically sensitive to polarization because it exploits Compton scattering\@.

The entire set of detectors and their cryostat are enclosed inside a well of anticoincidence BGO shields to reduce the Earth albedo and atmospheric backgrounds\@. The resulting overall field of view is primarily limited by the BGO shields to $\sim$3.2~sr\@. The entire instrument and readout electronics are mounted in a pointed, autonomous balloon gondola\@. Further details of the gondola systems can be found in~\cite{Bellm:2009:p1250}\@.

Data from the detectors are processed on-board by conventional surface-mount electronics \citep{Coburn:2004:p131,Bandstra:2007:p2532}, which return the energy and timing of each strip above threshold during an event\@.  The event information is processed on-board and telemetered to the ground\@. More details on the digital system, including flow charts of the digital logic involved in acquiring events, can be found in~\cite{Hung:2009:p2303}\@.

The software analysis tools for NCT are built using the Medium Energy Gamma-ray Astronomy library (MEGAlib)~\citep{Zoglauer:2006:p629,Zoglauer:2008:p101}\@. MEGAlib provides utilities for simulations~\citep{Zoglauer:2009:IEEECosima}, geometry modeling, event reconstruction, and imaging reconstruction~\citep{Zoglauer:2010:Mimrec}, in addition to tools for building a custom analysis pipeline\@.  The NCT data analysis pipeline is explained in further detail by~\cite{Bandstra:2009:p2131}\@.

\section{Data Selections}

\label{sec:data_selections}

NCT was launched from the Columbia Scientific Balloon Facility (CSBF) in Fort Sumner, New Mexico (34.5$^{\circ}$N, 104.2$^{\circ}$W) at 1330 UT 5/17/2009\@. The flight was terminated at 0400 UT 5/19/2009 near Kingman, Arizona (34.9$^{\circ}$N, 113.7$^{\circ}$W)\@. Altitude was maintained between 35 and 40~km for the entire time the detectors were operational\@. The total flight time was 38.5~hours, with nine of the ten detectors on for a total of 22~hours at float\@. Approximately 33~ks of data were taken with the Crab above 40$^{\circ}$ elevation\@. A discussion of the flight can be found in~\cite{Bandstra:2009:p2131}\@.

Data from the flight were selected only when the Crab was above 40$^{\circ}$ elevation both to minimize atmospheric absorption and to sample the background (as will be discussed in Sections~\ref{sec:analysis_spectrum} and \ref{sec:analysis_arm})\@. During the chosen observation time, the average livetime for the instrument was nearly constant with an average value of 92\%\@. Taking into account the livetime and $\approx$1~ks of dead time due to data acquisition system and flight computer resets, the effective Crab observation time was 29.3~ks\@.

\subsection{Event Reconstruction}
Event reconstruction for events with three or more interactions was done using the Compton Kinematic Discrimination (CKD) algorithm of \cite{Boggs:2000:p311}\@.  For events with $N$ interactions, there are $N-2$ Compton scatter angles that can be measured (i.e., at all but the first and last interaction points), and this allows for a figure-of-merit with $N-2$ degrees of freedom to be calculated for each possible event ordering\@.  The best interaction ordering of the $N!$ possible orderings is selected\@.  In principle, the more interactions there are, the more information there is about the event and the better it can be reconstructed (or rejected) using CKD\@.  A limit of seven interactions was chosen because the number of events with more than seven interactions becomes very small (fewer than 1\%) and the time to reconstruct a single event scales like $N$~factorial.

Two-interaction events (i.e.,~a Compton scatter and photoabsorption) lack any kinematic degrees of freedom and cannot be reconstructed using CKD.  Their interaction ordering is often ambiguous, and a single-scatter reconstruction technique must be used\@.  After trying different techniques on the NCT calibration data, we found the technique that properly reconstructed the most events was to choose the ordering that maximized the product of the Klein-Nishina cross section for the implied scatter angle with the probability of a photoabsorption at or before the second interaction\@.

\subsection{Background Reduction}
The gamma-ray background at balloon altitudes was measured by the NCT prototype during its 2005 balloon flight~\citep{Bowen:2009:Thesis}\@.  The background primarily consists of three kinds of background events: real photons coming from outer space and from cosmic-ray interactions in the atmosphere; cosmogenic particles such as protons and neutrons; and instrumental events coming from radioactivities in the instrument\@.

Simulations are used to estimate how many instrumental events survive the event reconstruction process\@.  Such comparisons to simulations have been done, and the result is that instrumental events are effectively eliminated by CKD, and the background is dominated by real photons -- predominantly atmospheric continuum photons, cosmic photons, and atmospheric 511 keV photons; and particle interactions are very few~\citep{Bowen:2009:Thesis}\@. The 2005 measurements yielded good agreement with simulations, and these simulation inputs were used again to estimate the background during the 2009 flight\@.

The atmospheric background continuum has been extensively studied and measured (e.g., \cite{Ling:1975:p3241, Ling:1977:p1211, Graser:1977:p1055, Schonfelder:1977:p306})\@.  The continuum has an elevation angle dependence --- the specific intensity should be approximately constant and then rise sharply at the horizontal\@.  The intensity in the horizontal direction should be larger than that from the vertical by a factor of $\approx$2--4 over the energy range 300--1000~keV\@.  This sharp rise in the horizontal direction is another reason for requiring that the Crab be above 40$^{\circ}$ in elevation\@.

In addition, the NCT's BGO shielding was designed so that it would have a very wide field of view away from the Earth, reducing background photons that are coming from the atmosphere below and from the sides\@.  An Earth horizon cut (Table~\ref{tab:data_cuts}) is also applied to further reduce background that is coming from the atmosphere below\@. However, background photons from the atmosphere above can only be reduced by shielding, with a corresponding reduced field of view\@.

An energy cut of 300$-$495~keV and 520$-$1500~keV excluded a few other sources of background\@. The cut avoids the very strong atmospheric 511~keV background line\@. Also, below $\approx$300~keV, discrepancies were noticed between the observed background continuum and simulations, with NCT observing fewer counts by a factor of two in that region~\citep{Chiu:2010:Thesis}\@. The comparison of simulations and effective area calibration data at higher energies reveals a deficit of measured photons when a source is placed in the plane of the detectors, presumably because the simulations underestimate the shielding effect of the detector guard rings\@. This effect is more pronounced for lower energy photons and likely leads to the measured deficit\@. This low-energy region ($<$300~keV) will be excluded from the analysis\@.

\subsection{Interaction Distance}
It also became clear from simulations and measured background rates that the other data cuts used for the flight data should be wide in order to attain the best sensitivity\@. We chose one major constraint --- to limit the distance between the first and second interactions in the detectors --- in order to improve the angular resolution of the instrument and thus the signal-to-noise\@. This choice is effective because the position resolution of the NCT detectors is the the dominant contribution to the angular resolution~\citep{Bandstra:2006:p770}, and the vector between the first two interactions sets the axis of the Compton circle\@.  Therefore, requiring a longer initial scatter distance improves the angular resolution, but the cost is that the number of events is reduced, especially those with three or more interactions\@.  Based on simulations of the Crab, setting a lower limit of 1.2~cm for the first interaction distance improves the angular width of the Crab (10$^{\circ}$~ARM~FWHM to 7.4$^{\circ}$~ARM~FWHM), while eliminating only 40\% of the events within the central FWHM of the ARM distribution (see Section~\ref{sec:arm_def} for a definition and discussion of the ARM)\@. The cut is also meant to improve the likelihood that an event is properly reconstructed by ensuring that the first two interactions are separated by multiple strips\@.

The different cuts used in the analysis of the NCT flight data are summarized in Table~\ref{tab:data_cuts}\@.
After applying these data selections, 65.8\% of the remaining events had two interactions, 25.9\% had three interactions, 6.6\% had four interactions, and 1.7\% had five or more\@.

\begin{table}
	\begin{center}
		\caption{Data cuts used on flight data\label{tab:data_cuts}}
		\begin{tabular}{>{\centering}p{4cm}|>{\centering}p{4cm}}
			\hline
			\hline
			\textbf{Parameter} & \textbf{Allowed range}\tabularnewline
			\hline
			Crab elevation & $>$40$^{\circ}$\tabularnewline
			\hline
			Photon energy & 300$-$495 keV, 520$-$1500 keV\tabularnewline
			\hline
			Number of interactions & 2$-$7\tabularnewline
			\hline
			Earth horizon & Reject if $>$99\% of Compton circle is below elevation of 0$^{\circ}$\tabularnewline
			\hline
			Compton scatter angle ($\varphi$) & no restriction (0$^{\circ}$$-$180$^{\circ}$)\tabularnewline
			\hline
			Minimum distance between first two interactions & 1.2 cm\tabularnewline
			\hline
			Minimum distance between any two interactions & 0.4 cm\tabularnewline
			\hline
		\end{tabular}
	\end{center}
\end{table}

\section{Image of the Crab Nebula}

\label{sec:image}

After applying the data cuts, an image was produced from the remaining data\@. Since NCT measures the location of the photon to a circle on the sky (not a point), Compton imaging techniques are used to produce images\@. Compton imaging can be done through the straightforward backprojection of Compton circles onto the sky or with more sophisticated image processing\@. One imaging tool that has been used for NCT data is List-Mode Maximum Likelihood Expectation Maximization (MLEM)~\citep{Wilderman:1998:p1716}, which has been implemented in MEGAlib~\citep{Zoglauer:2010:Mimrec}\@. This method is iterative, and it uses a maximum likelihood statistic to refine the raw backprojection\@. The result is that point sources become clearer and other features are sharpened\@.

\begin{figure*}
	\begin{center}
		\ifthenelse{\boolean{electronic_on}}
		{\plotone{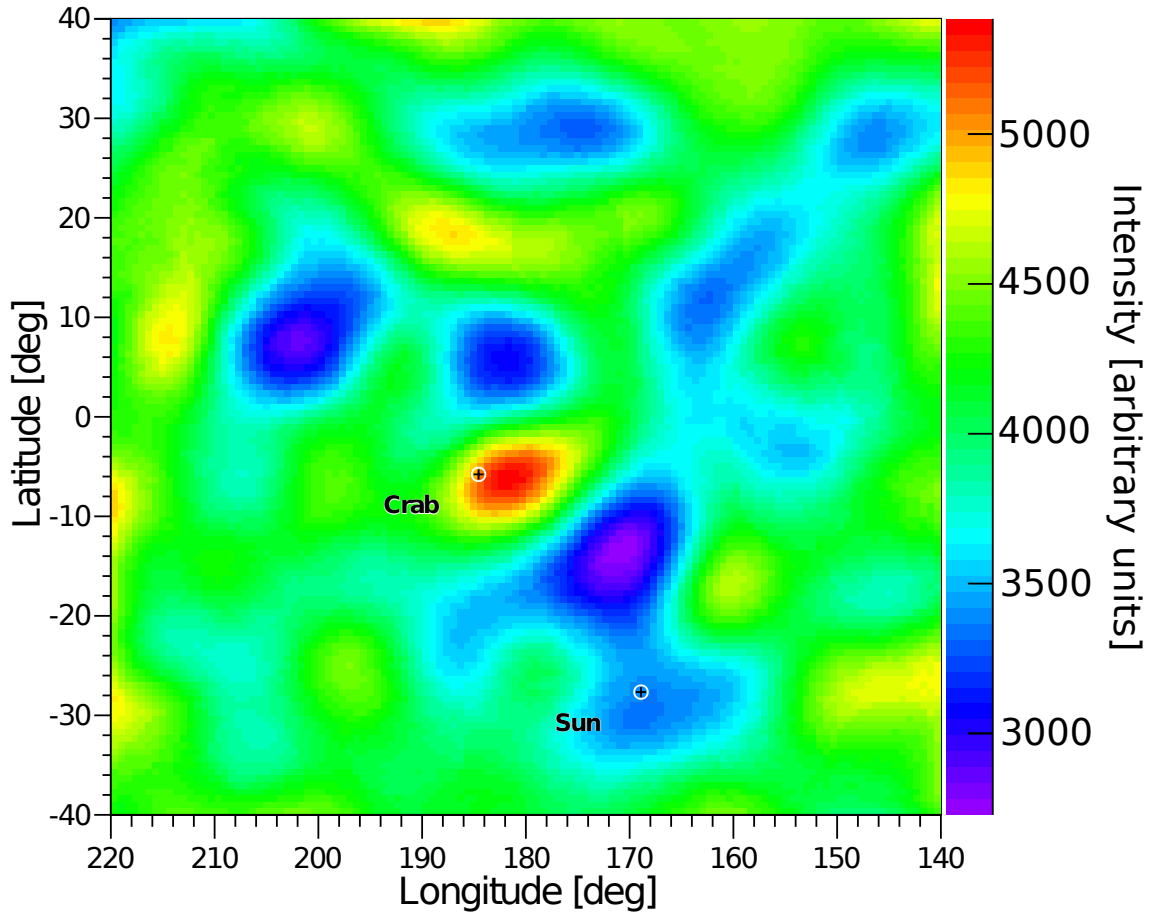}}
		{\plotone{f1.eps}}
	\end{center}
	\caption{Image of the NCT flight data in the region of the Crab, with the positions
	of the Crab and the Sun marked\@.  Five iterations of the
	MLEM algorithm have been performed\@. There is a clear enhancement in the vicinity of the Crab\@.
	The color scale is an arbitrary intensity scale\@.
	\label{fig:image_iter}}
\end{figure*}

\begin{figure*}
	\begin{center}
		\ifthenelse{\boolean{electronic_on}}
		{\plotone{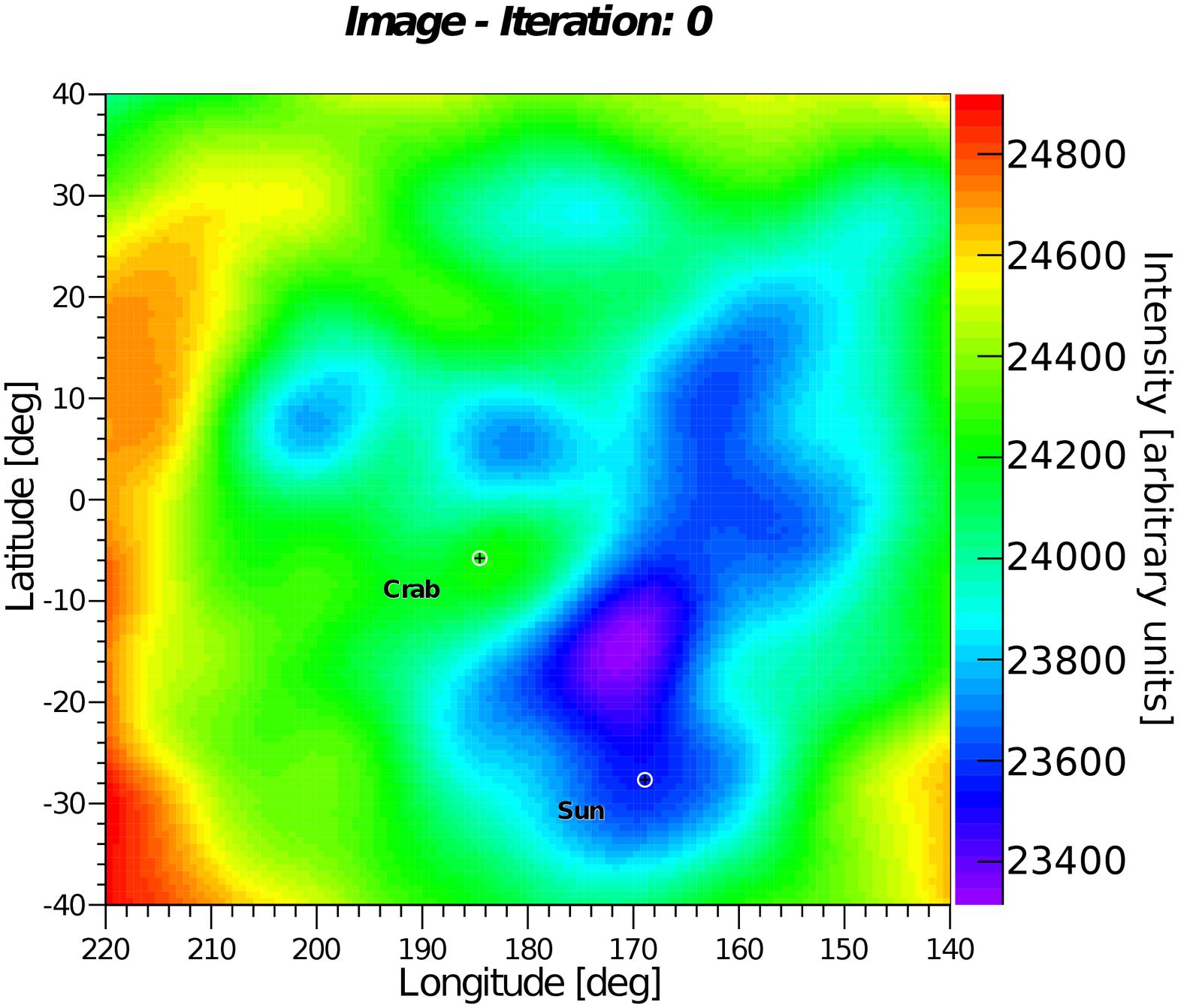}}
		{\plotone{f2.eps}}
	\end{center}
	\caption{Raw Compton circle backprojection of NCT flight data in galactic coordinates, with the
	position of the Crab and the Sun marked\@. A total of 289,128 event
	circles are plotted, and $\approx 1.3$\% of that number are expected to come from the Crab\@.
	Notice that there is no clear source at the location of the Crab before image processing\@.
	The color scale is an arbitrary intensity scale\@.	
	\label{fig:image_raw}}
\end{figure*}

We constructed an image from the NCT flight data using the cuts mentioned previously\@. We plotted a total of 289,128 event circles in a $80^{\circ}\times80^{\circ}$ backprojection, of which approximately 3,800 event circles (1.3\%) are expected from the Crab based on simulations (see Section~\ref{sec:sim})\@. We performed a total of five iterations of the MLEM algorithm on the image, which is the approximate number of iterations needed so that the FWHM of a point source image matches the 7.4$^{\circ}$~ARM~FWHM\@.  The result is that the Crab appears in the image (Fig.~\ref{fig:image_iter}), though it was not originally apparent in the backprojection (Fig.~\ref{fig:image_raw})\@. The Sun was in the field of view but no source appears there, as expected\@.

The Crab image reveals a slight offset ($\approx2^{\circ}$) in the position of the source\@. The significance of this offset can be evaluated in the following manner\@. Given the equivalent gaussian sigma of the ARM histogram for the Crab ($7.4^{\circ}/2.35=3.1^{\circ}$) and the signal-to-noise estimate of $4\sigma$ (Section~\ref{sec:significance}), the one-sigma uncertainty of the image position can be estimated as $\approx3.1^{\circ}/4=0.8^{\circ}$ (e.g.,~\cite{Bobroff:1986:p1152})\@. This statistical uncertainty, combined with the $\approx1^{\circ}$ systematic uncertainty in the gondola aspect, means the $\approx2^{\circ}$ offset in the source position is not statistically significant\@.

\section{Simulation of the Crab Observation}

\label{sec:sim}

A simulation of the Crab Nebula spectrum was performed as a point of comparison for the NCT flight data\@. A detailed mass model for NCT has been developed, along with code that applies the detector response to the ideal simulated detector hits~\citep{Bandstra:2009:p2131, Chiu:2009:IEEE}\@. The MEGAlib simulation tool Cosima was used, which is based on the gamma-ray telescope library MEGAlib and the Monte Carlo framework GEANT4~\citep{Agostinelli:2003:p250}\@. A point source with a spectrum identical to the Crab Nebula spectrum measured by INTEGRAL/SPI~\citep{Jourdain:2009:p17} was simulated at infinity along the z-axis of the detectors (i.e.,~the center of the field of view) for an equivalent real time of 100,000~seconds\@. The total Crab flux between 200~keV and 10~MeV was calculated to be $2.194\times10^{-2}$~ph~cm$^{-2}$~s$^{-1}$\@. The total number of events between 200~keV and 10~MeV that interacted in the detectors were 219,987, with 18,622 events passing the event cuts (8.5\%)\@.

Compensation for atmospheric absorption was done after the simulation by calculating the transmission probability as a function of energy and multiplying the probability with the spectrum of events that passed the data cuts\@. The more accurate way of accounting for atmospheric absorption is to perform Monte Carlo simulations with a layer of air of the appropriate column density, but a brief exploration of such simulations revealed that there was little difference\@. An average column density of 3.2~g~cm$^{-2}$ was calculated for the chosen observation time, and the Crab spectrum at 45$^{\circ}$, 55$^{\circ}$, 65$^{\circ}$, and 75$^{\circ}$ was calculated (the Crab elevation peaks at 80$^{\circ}$ at transit)\@. An average spectrum was calculated, and it contained 71\% of the total counts of the original spectrum\@. This transmission percentage was used as an approximation for atmospheric absorption in the simulations\@.

\section{The Angular Resolution Measure (ARM)}

\label{sec:arm_def}

\begin{figure}
	\begin{center}
		\ifthenelse{\boolean{aastex_on}}
		{\epsscale{0.5}}{}
		\plotone{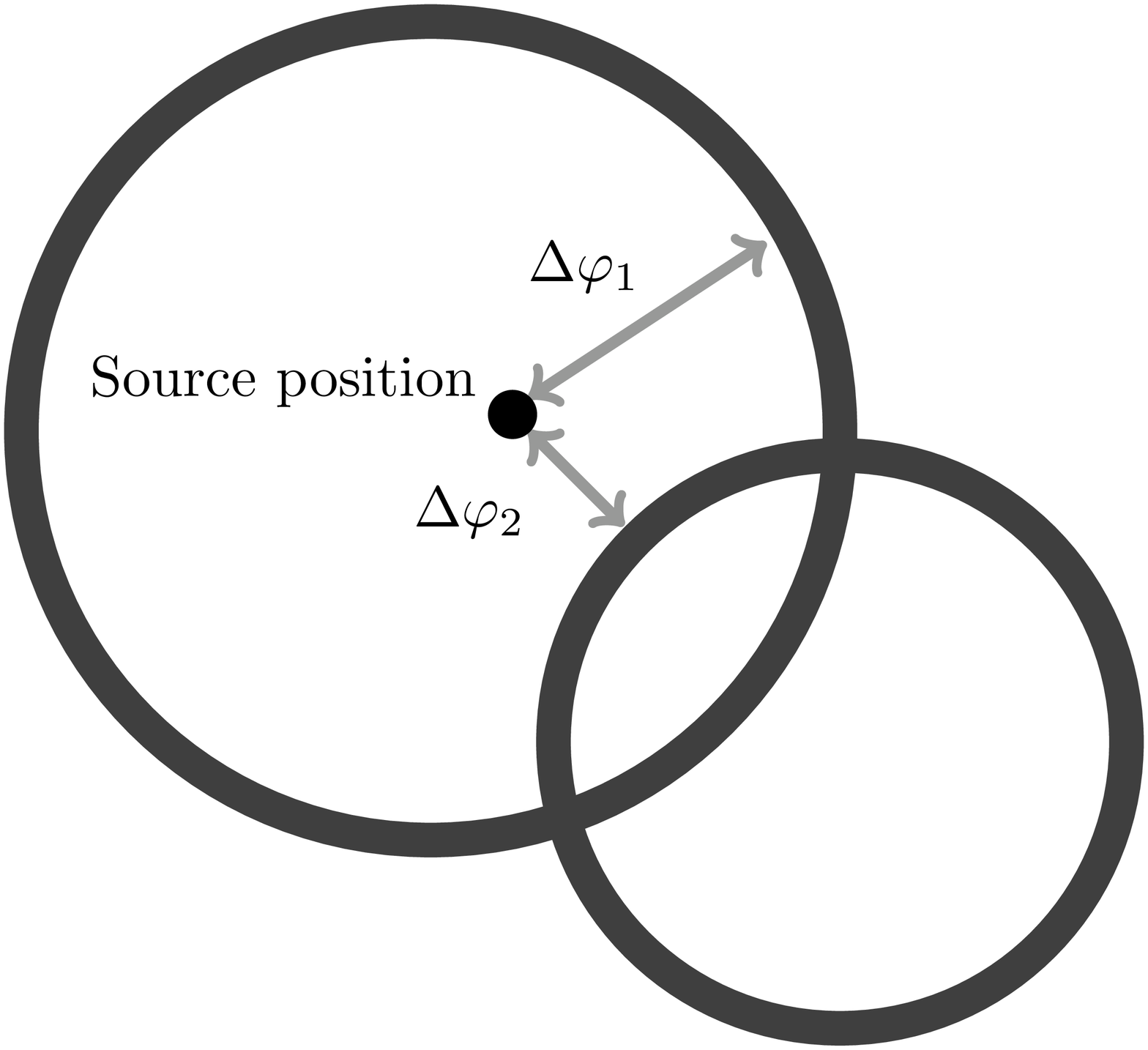}
	\end{center}
	\caption{Schematic of the definition of the Angular Resolution Measure (ARM)\@.
	Notice that the ARM can be positive or negative: $\Delta\varphi_{1}<0$,
	since its reconstructed Compton circle circumscribes the source position,
	while $\Delta\varphi_{2}>0$ because its Compton circle does not circumscribe
	the source position\@.\label{fig:ARM_definition}}
\end{figure}

An important tool that will be used in the data analysis is the Compton telescope Angular Resolution Measure (ARM)\@. The ARM is a statistic that is used to measure the angular resolution of a Compton telescope (e.g.,~\cite{Schonfelder:1993:p657})\@. An ARM histogram can be constructed when the source's location is known\@. The ARM histogram is a histogram of the following quantity:
\begin{equation}
\Delta\varphi_{\rm{ARM}}\equiv\varphi_{\rm{geo}}-\varphi_{\rm{recon}}\label{eq:ARM_def-1}
\end{equation}
where $\varphi_{\rm{geo}}$ is the scatter angle derived from the known source location and $\varphi_{\rm{recon}}$ is the Compton scatter angle obtained from event reconstruction (i.e.,~the Compton scatter formula)\@. For a single event, $\Delta\varphi_{\rm{ARM}}$ measures the deviation of the event's Compton circle from the true position of the source\@. A graphical explanation of this definition is shown in Fig.~\ref{fig:ARM_definition}\@. The angular resolution of a Compton telescope is specified by the full-width at half maximum (FWHM) of the ARM histogram\@.

The FWHM of the ARM histogram is affected by detector energy resolution, detector position resolution, and Doppler broadening\@. Doppler broadening is the deviation from the ideal Compton scatter formula caused by scattering off of bound atomic electrons (not free electrons)\@. It is a fundamental limit to the angular resolution of a Compton telescope\@. This broadening has been extensively studied for different materials in~\cite{Zoglauer:2003:p1302}\@. Because of NCT's excellent energy resolution, the contribution from the energy resolution is not important\@. The position resolution of the NCT detectors (especially the 2~mm strip pitch) is the the dominant contribution to the ARM~\citep{Bandstra:2006:p770}\@. This is one reason the event cut on the first interaction distance greatly improves the ARM FWHM\@.

The ARM is dominated by the distance between the first two interactions, because that vector forms the axis of the Compton circle\@. Subsequent scatters do not affect the ARM greatly; if anything, events with multiple Compton scatters have a narrower ARM\@. This can be understood by noting that in order to undergo further scatters, the photon must lose less energy in the first scatter\@. Thus the Compton scatter angle for these events must be on average smaller, reducing the uncertainty in the ARM\@.

Since the position resolution of all the detectors is the same (2~mm GeD strip pitch, and 0.5~mm depth resolution), and the GeD energy resolutions are similar ($~$2--4~keV~FWHM at 662~keV), the ARM is uniform throughout the detectors. The ARM is predominantly affected by the geometry of the event.  In fact, one can estimate NCT's ARM FWHM by using the requirement of at least 1.2~cm between the first two scatters and the germanium detector strip pitch of 2~mm\@. The FWHM should be $~\frac{0.2}{1.2} = $9.5$^{\circ}$, which is comparable to the 12.8$^{\circ}$ measured for the 662 keV calibration source and the 7.4$^{\circ}$ measured for the Crab itself\@.  By comparison, the contribution to the ARM due to the energy resolution (assuming $\approx$3~keV~FWHM) should be $~\frac{511\cdot 3}{662^2} = $0.3$^{\circ}$\@.

The ARM will be used to select events from the location of the Crab and test points\@. The ARM distribution can usually be approximated by a Lorentzian distribution with FWHM $\Gamma$, and the sensitivity of data cuts is maximized by selecting events that are within $0.7\Gamma$ of the center\@. For the data cuts we are using, $\Gamma=7.4^{\circ}$, so we use an ARM cut of $\pm5.2^{\circ}$ to maximize sensitivity\@.

\section{Significance of the Crab detection}

\label{sec:significance}

To measure the significance of the Crab, events that were consistent with the Crab were extracted from the data\@.  Events whose Compton circles passed within the optimal ARM cut of $5.2^{\circ}$ around the position of the Crab were selected\@.

An estimate of the background counts is needed to determine the significance\@. Because the Crab moves through NCT's wide field of view, we can select nearby points to estimate the background at the Crab's position\@. An added complication is that because photons are only localized to a circle, a photon from an off-source point may be consistent with the Crab as well\@. For this analysis, we selected eight points that were on a circle $\delta=20{}^{\circ}$ away from the Crab at evenly-spaced intervals\@. Events were selected from within $5.2^{\circ}$ of these points\@. The angular separation of $\delta=20^{\circ}$ was selected so that we could sample points that were close enough to the Crab that they have similar exposure times, and far enough away so that fewer Compton circles would be consistent with both the Crab and these points\@. There is a trade-off between choosing an offset angle $\delta$ so that the background estimate is reliable (i.e.,~$\delta$ should be small) and so that the off-axis source contribution is small (i.e.,~$\delta$ should be large)\@. The value of 20$^{\circ}$ was chosen in this work as the largest angle such that the interval $(-\delta,+\delta)$ could be sampled for the later ARM analysis (due to the 40$^{\circ}$ minimum elevation of the Crab)\@.

Table~\ref{tab:total_counts} shows the total number of events that were extracted from the on-source and the off-source positions\@. The results of the same analysis on the Crab simulations (i.e.,~no background) are also shown in the table\@. The number of excess counts observed at the position of the Crab agrees with the simulations to within statistical error\@.

\begin{table}
	\begin{center}
		\caption{The number of events with Compton circles within 5.2$^{\circ}$ of the Crab\label{tab:total_counts}}
		\begin{tabular}{>{\centering}m{1.5cm}|>{\centering}p{2.5cm}|>{\centering}p{3.0cm}}
			\hline
			\hline 
			Position & Counts (Flight Data) & Counts (Scaled Simulation of Crab)\tabularnewline
			\hline
			$(\ell_{0},b_{0})$ & 29,808 & 1101\tabularnewline
			\hline 
			Eight sample points 20$^{\circ}$ away & \begin{tabular}{c}
			29,041\tabularnewline
			28,708\tabularnewline
			29,453\tabularnewline
			29,462\tabularnewline
			29,124\tabularnewline
			28,917\tabularnewline
			29,093\tabularnewline
			29,329\tabularnewline
			\end{tabular} & \begin{tabular}{c}
			481\tabularnewline
			459\tabularnewline
			444\tabularnewline
			453\tabularnewline
			482\tabularnewline
			457\tabularnewline
			432\tabularnewline
			447\tabularnewline
			\end{tabular}\tabularnewline
			\hline 
			Average off-source & 29,141 & 457\tabularnewline
			\hline 
			Excess on-source & 667 & 644\tabularnewline
			\hline
		\end{tabular}
		\tablecomments{The point $(\ell_{0},b_{0})=(184.56^{\circ},-5.78^{\circ})$ is the position of the Crab in galactic coordinates\@.}
	\end{center}
\end{table}

The significance of the excess source counts can be estimated in the following way\@. The na\"{i}ve approach is to assume that all of the measurements (off-source and on-source) are independent Poisson processes\@. Let the Crab point be labeled as $C$, and the off-source points labeled as $i=1\dots8$\@. Then this approach yields an approximate statistical uncertainty on the excess counts:
\begin{eqnarray}
N_{{\rm excess}} & = & N_{C}-\bar{N}_{{\rm off-source}} = N_{C}-\frac{1}{8}\sum_{i=1}^{8}N_{i} \\
\Rightarrow\hat{\sigma}_{{\rm excess}}^{2} & = & \sigma_{N_{C}}^{2}+\frac{1}{64}\sum_{i=1}^{8}\sigma_{N_{i}}^{2} = N_{C}+\frac{1}{64}\sum_{i=1}^{8}N_{i}\\
\hat{\sigma}_{{\rm excess}} & = & 182.9
\end{eqnarray}
This approach is not exact because the Compton event circles are likely to intersect multiple points, meaning $N_{C}$ and $N_{i}$ are not independent Poisson measurements\@. We identify the following independent Poisson processes that can be used to calculate a valid statistical uncertainty:
\begin{equation}
\begin{array}{ll}
\bar{C}+k: & \mbox{An event intersects \ensuremath{k\:}off-source points}\\
 & \mbox{ \textbf{and not} the Crab point}\\
C+k: & \mbox{An event intersects \ensuremath{k\:}off-source points}\\
 & \mbox{ \textbf{and} the Crab point}\end{array}
\end{equation}
where $k=0\dots8$\@. Let $N_{\bar{C}+k}$ and $N_{C+k}$ be the counts measured for each Poisson process\@. These measurements are listed in Table~\ref{tab:snr_calc}\@. Then the number of excess Crab counts can be written in the following way:
\begin{eqnarray}
N_{{\rm excess}} & = & N_{C}-\bar{N}_{{\rm off-source}}\\
 & = & N_{C}-\frac{1}{8}\sum_{i=1}^{8}N_{i}\\
 & = & \sum_{k=0}^{8}N_{C+k}-\frac{1}{8}\left(\sum_{k=1}^{8}kN_{C+k}+\sum_{k=1}^{8}kN_{\bar{C}+k}\right)\\
 & = & N_{C+0}+\sum_{k=1}^{8}\left(1-\frac{k}{8}\right)N_{C+k}-\frac{1}{8}\sum_{k=1}^{8}kN_{\bar{C}+k}.
\end{eqnarray}
The factors of $k$ in the off-source sum are needed because each event tallied by $N_{C+k}$ or $N_{\bar{C}+k}$ is counted at $k$ different off-source points\@. The statistical uncertainty for the excess counts becomes
\begin{eqnarray}
\sigma_{{\rm excess}}^{2} & = & N_{C+0}+\sum_{k=1}^{8}\left(1-\frac{k}{8}\right)^{2}N_{C+k}+\sum_{k=1}^{8}\left(\frac{k}{8}\right)^{2}N_{\bar{C}+k}\\
\Rightarrow\sigma_{{\rm excess}} & = & 163.4
\end{eqnarray}
The result of this calculation is that the Crab was detected at the $4.1\sigma$ level, which is within statistical error of the expected $3.9\sigma$ from simulations\@.

\begin{table}
	\begin{center}
		\caption{Data used in the calculation of the signal-to-noise ratio of the Crab data \label{tab:snr_calc}}
		\begin{tabular}{>{\centering}p{2.4cm}|>{\centering}p{2.cm}|>{\centering}p{2.cm}}
			\hline
			\hline
			$k$ & $N_{\bar{C}+k}$ & $N_{C+k}$\tabularnewline
			\hline
			0 & N/A & 2578\tabularnewline
			\hline
			1 & 29,209 & 12,801\tabularnewline
			2 & 61,448 & 14,373\tabularnewline
			3 & 11,752 & 51\tabularnewline
			4 & 803 & 5\tabularnewline
			5 & 88 & 0\tabularnewline
			6 & 27 & 0\tabularnewline
			7 & 8 & 0\tabularnewline
			8 & 22 & 0\tabularnewline
			Total & 103,357 & 29,808\tabularnewline
			\hline 
			\hline 
 			& Flight: & Simulation:\tabularnewline
			\hline 
			Excess on-source & 667 & 644\tabularnewline
			\hline 
			Background noise ($\sigma$) & 163.4 & 163.4 (using flight value)\tabularnewline
			\hline 
			\textbf{Significance} & \textbf{4.1$\boldsymbol{\sigma}$} & \textbf{3.9$\boldsymbol{\sigma}$}\tabularnewline
			\hline
		\end{tabular}
	\end{center}
\end{table}

\section{Analysis of the Crab spectrum}

\label{sec:analysis_spectrum}

\begin{figure}
	\begin{center}
		\ifthenelse{\boolean{aastex_on}}
		{\epsscale{0.7}}{}
		\plotone{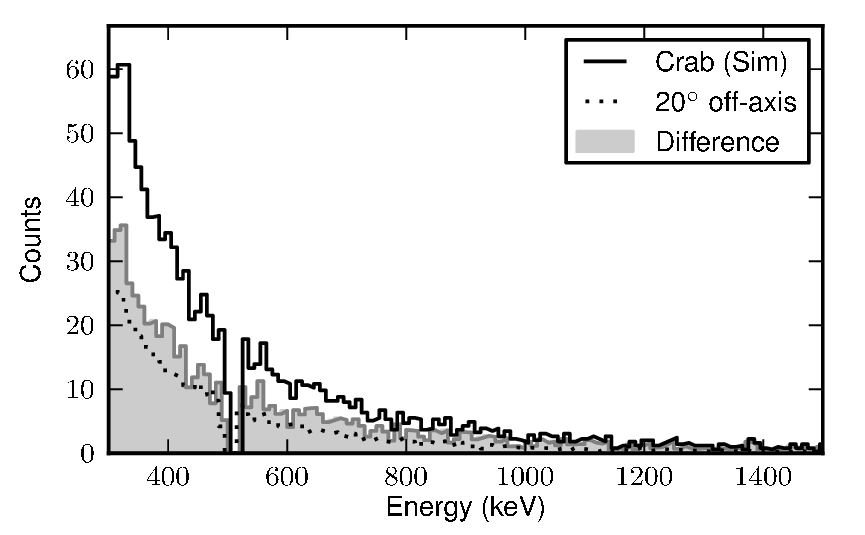}
	\end{center}
	\caption{Simulated spectrum for events within 5.2$^{\circ}$ of the Crab using
	data cuts given in the text\@. The averaged spectrum measured
	at points 20$^{\circ}$ away is also plotted\@. The difference of
	these two distributions is the signature we expect in the flight
	data after estimating the spectrum at nearby points\@.\label{fig:spec_sim_offaxis}}
\end{figure}

To examine the spectrum of the Crab, a spectrum is formed for the on-source data, and an average off-source spectrum is also calculated\@. Error bars are calculated for each spectral bin in the same manner described in Section~\ref{sec:significance} and Table~\ref{tab:snr_calc}\@. The results of this procedure for the simulated Crab alone are shown in Fig.~\ref{fig:spec_sim_offaxis}\@.

The chi-squared statistic is calculated for the measurement assuming there is no source present ($\chi^2_{0}$) and the measurement assuming the Crab spectrum as derived from simulations ($\chi^2_{\rm{Crab}}$)\@. When combined with the number of degrees of freedom of the data ($\nu$), each chi-squared statistic yields a $P$-value, or the probability that random fluctuations alone would yield a chi-squared of equal or greater value\@.  The $P$-value for the null hypothesis quantifies whether a source is present ($P_0 \ll 1$) or not ($P_0 \sim 1$), and the $P$-value for the Crab quantifies whether the measured spectrum is statistically consistent with the simulation ($P_{\rm{Crab}} \sim 1$) or not ($P_{\rm{Crab}} \ll 1$)\@. An equivalent gaussian sigma $n$ can be derived from a $P$-value by equating the chi-squared $P$-value with the complementary cumulative distribution function of a gaussian random variable: $n = \sqrt{2}\ \mbox{erf}^{-1}\left(1-P\right)$\@.

\begin{figure}
	\begin{center}
		\ifthenelse{\boolean{aastex_on}}
		{\epsscale{0.7}}{}
		\plotone{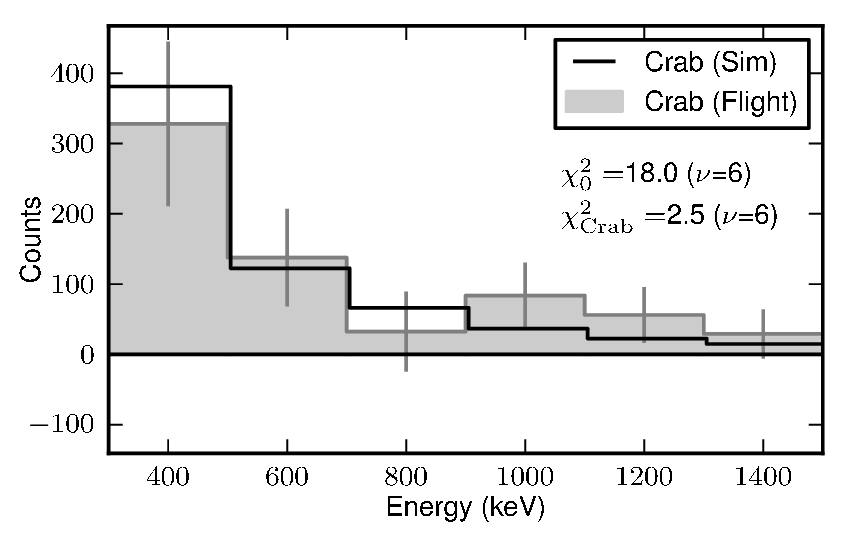}
	\end{center}
	\caption{Measured spectrum for the Crab plotted with the expected excess
	from simulations\@. The spectrum is measured by subtracting the
	average spectrum from points 20$^{\circ}$ away from the Crab\@. Error
	bars are calculated using Poisson statistics\@. The measured spectrum
	is inconsistent with zero ($2.7\sigma$) but is consistent with the
	simulated Crab spectrum ($P_{\rm{Crab}}=0.87$)\@. The bins are 200~keV wide\@.\label{fig:spec_flight}}
\end{figure}

\begin{figure}
	\begin{center}
		\ifthenelse{\boolean{aastex_on}}
		{\epsscale{0.5}}{}
		\plotone{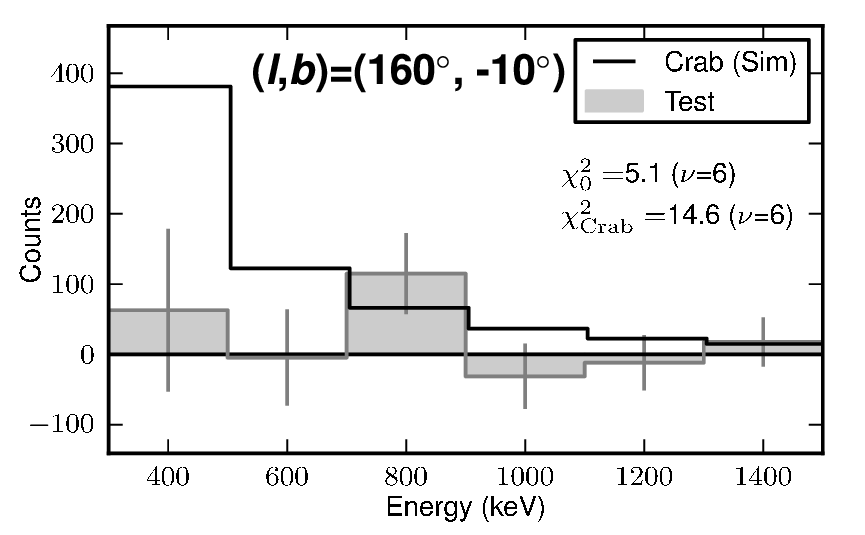}\\
		\plotone{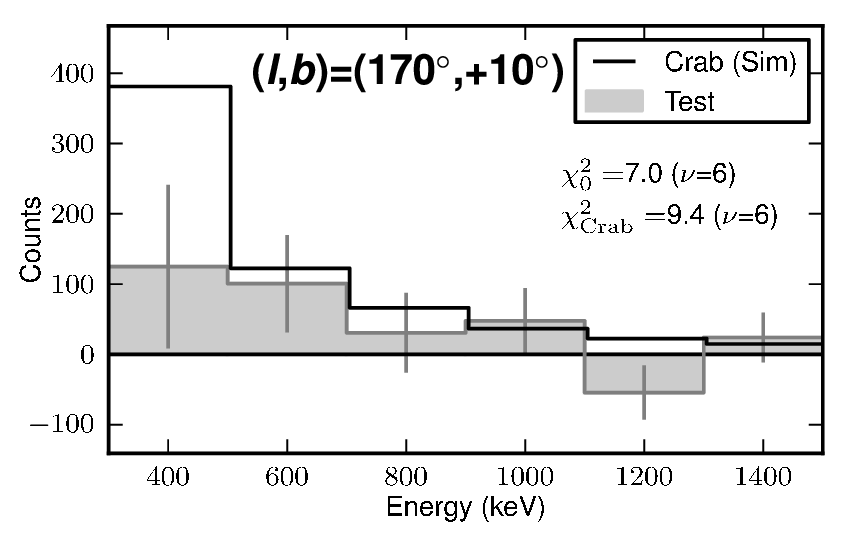}\\
		\plotone{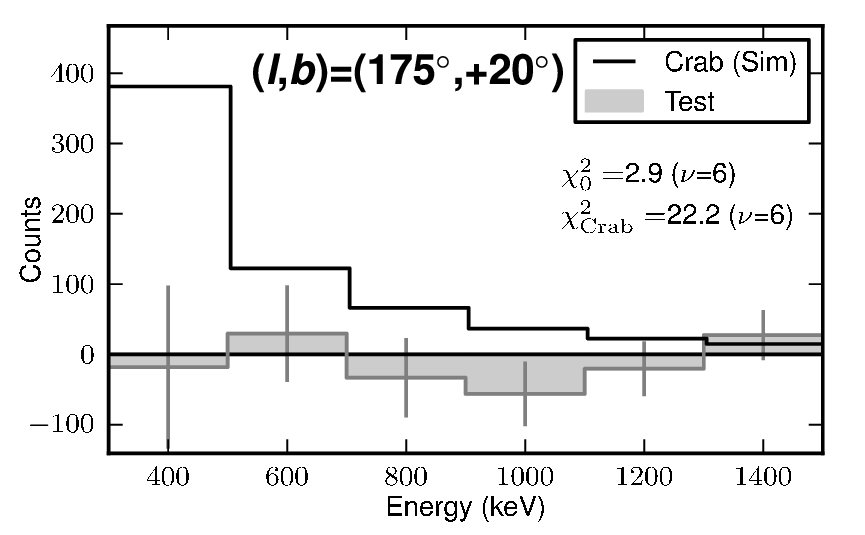}
	\end{center}
	\caption{Measured spectrum at three different test points, with
	the expected excess for the actual Crab position shown\@. The spectrum is
	measured by subtracting the average spectrum from eight points 20$^{\circ}$
	away from each test point\@. The spectrum at the three test points is
	consistent with zero\@.\label{fig:spec_flight_testpoints}}
\end{figure}

Performing this analysis on the flight data (Fig.~\ref{fig:spec_flight}) reveals a spectrum that is inconsistent with a null result with a $P$-value of $P_0=0.0062$ and $n_0 = 2.7\sigma$, and the spectrum is consistent ($P_{\rm{Crab}}=0.87$ and $n_{\rm{Crab}}=0.2\sigma$) with the simulations\@. To test the spectral analysis method, three other test points were chosen that had similar exposure during the flight: (160$^{\circ}$, --10$^{\circ}$), (170$^{\circ}$, +10$^{\circ}$) and (175$^{\circ}$, +20$^{\circ}$)\@. Each of these points yields spectrum consistent with the null hypothesis (the respective $n_0$ values are $0.6\sigma$, $1.0\sigma$, and $0.2\sigma$), and mostly inconsistent with the Crab spectrum ($n_0$ = $2.3\sigma$, $1.4\sigma$, and $3.3\sigma$)\@. The results are shown in Fig.~\ref{fig:spec_flight_testpoints}\@.

\section{Analysis of the Crab ARM}

\label{sec:analysis_arm}

The measured ARM histogram of the Crab should also match the simulated ARM histogram of the Crab in order to further confirm the detection\@. In order to see if there are excess counts in the ARM histogram at the location of the Crab, one must first estimate what the ARM should be in the absence of a source\@. To do this, we used the same eight $20^{\circ}$ off-axis points that were used in the previous analyses\@. As mentioned earlier, the value of $\delta=20^{\circ}$ was chosen in this work as the largest angle such that the full interval $(-\delta,+\delta)$ could be sampled in this analysis (due to the 40$^{\circ}$ minimum elevation of the Crab)\@. The ARM histogram was calculated for each of those points, and the average was constructed\@.

\begin{figure}
	\begin{center}
		\ifthenelse{\boolean{aastex_on}}
		{\epsscale{0.7}}{}
		\plotone{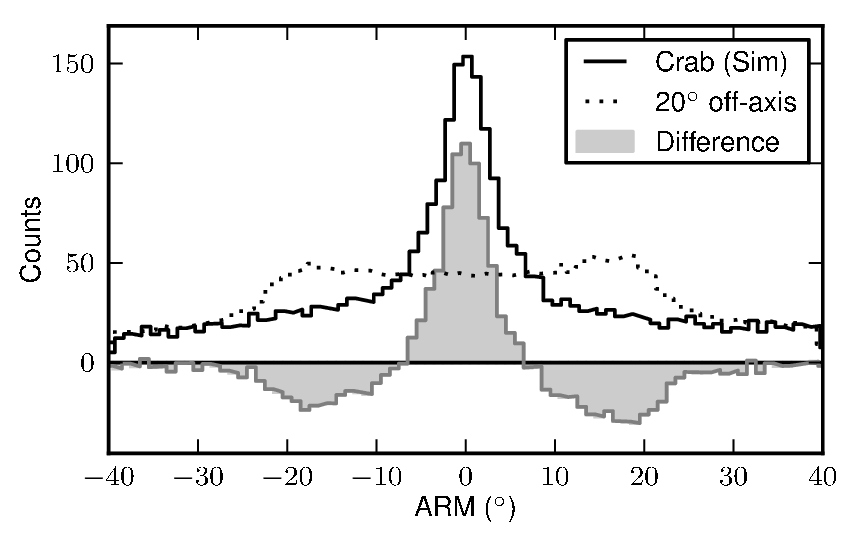}
	\end{center}
	\caption{Simulated ARM distribution for the Crab as seen by NCT using the data cuts given
	in the text\@. The FWHM of the simulated ARM distribution is 7.4$^{\circ}$~ARM~FWHM,
	as mentioned previously\@.
	The averaged ARM distribution at points 20$^{\circ}$
	away is also plotted\@. The difference of these two distributions
	is the signature we expect in the flight data after estimating
	the ARM at nearby points\@.\label{fig:arm_sim_offaxis}}
\end{figure}

An ARM histogram that is calculated at a point separated from a point source by an angle $\delta$ (20$^{\circ}$ in this case) will have an ARM histogram that no longer resembles a Lorentzian, but it is a mostly flat distribution between $\pm\delta$ that decays away beyond that region\@. So the points we used to estimate the ARM have photons from the Crab in them, and not an insignificant number\@. We must account for the stray counts from the Crab in some way or risk underestimating the flux\@. An example of this off-source ARM contribution is shown in Fig.~\ref{fig:arm_sim_offaxis}, where the on-source Crab simulation is shown with the 20$^{\circ}$ off-source Crab ARM\@. The difference between the on-source and off-source ARM histograms is the signature that we expect to see in the flight data when subtracting background estimates that are an angle $\delta$ away from the Crab\@.

\begin{figure}
	\begin{center}
		\ifthenelse{\boolean{aastex_on}}
		{\epsscale{0.7}}{}
		\plotone{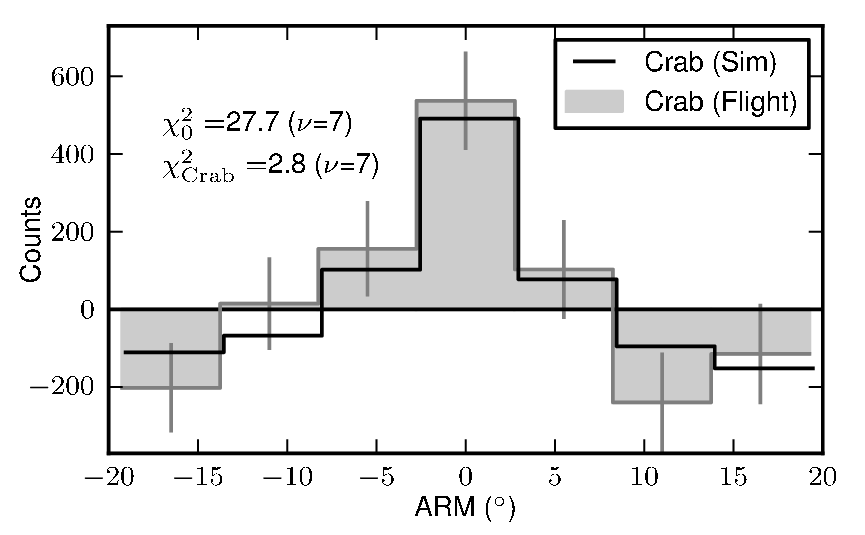}
	\end{center}
	\caption{Measured ARM excess for the Crab plotted, with the expected excess
	from Crab simulations shown\@. The spectrum is measured by subtracting
	the average spectrum from eight points 20$^{\circ}$ away from the
	Crab\@. Error bars are calculated using Poisson statistics\@. The measured
	ARM excess is significantly inconsistent with zero counts ($P_0=0.00025$,
	or $3.7\sigma$) and is compatible with the Crab simulation ($P_{\rm{Crab}}=0.90$)\@.
	The signal-to-noise ratio of the central bin is $4.2\sigma$\@. These
	results confirm the $4\sigma$ significance found in Section~\ref{sec:significance}\@.\label{fig:arm_flight}}
\end{figure}

The difference between the ARM histogram at the Crab position and the average ARM histogram of the off-source points was formed, and the error bars of each bin were calculated in the same manner described in Section~\ref{sec:significance} and Table~\ref{tab:snr_calc}\@. The result is shown in Fig.~\ref{fig:arm_flight}\@. The measured ARM excess is significant ($P_0=0.00025$, or $3.7\sigma$)\@. The ARM distribution is also very close to the simulated expectation ($P_{\rm{Crab}}=0.90$)\@. In addition, the central bin, which has the largest number of counts, has a signal-to-noise ratio of $4.2\sigma$, similar to the overall source significance derived in Section~\ref{sec:significance}\@. The expected simulation value for the central bin is $3.9\sigma$, so the simulations are in close agreement with the flight data\@.

\begin{figure}
	\begin{center}
		\ifthenelse{\boolean{aastex_on}}
		{\epsscale{0.5}}{}
		\plotone{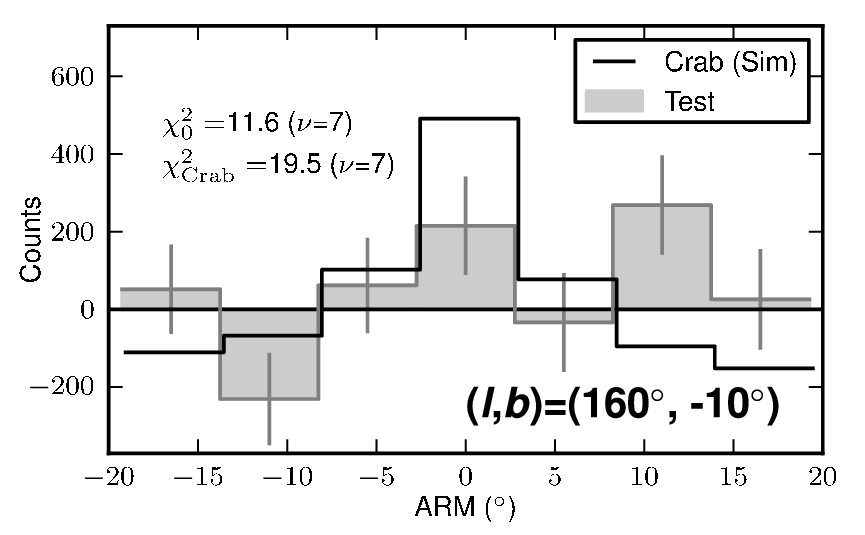}\\
		\plotone{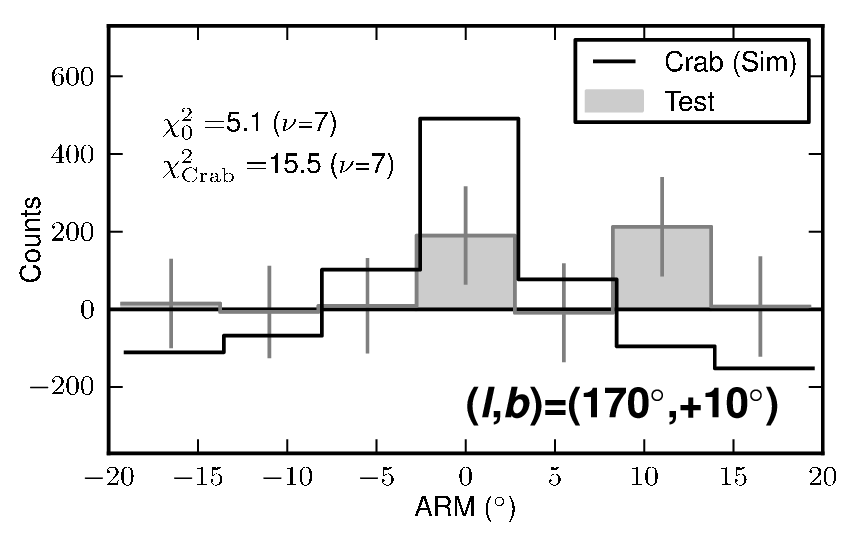}\\
		\plotone{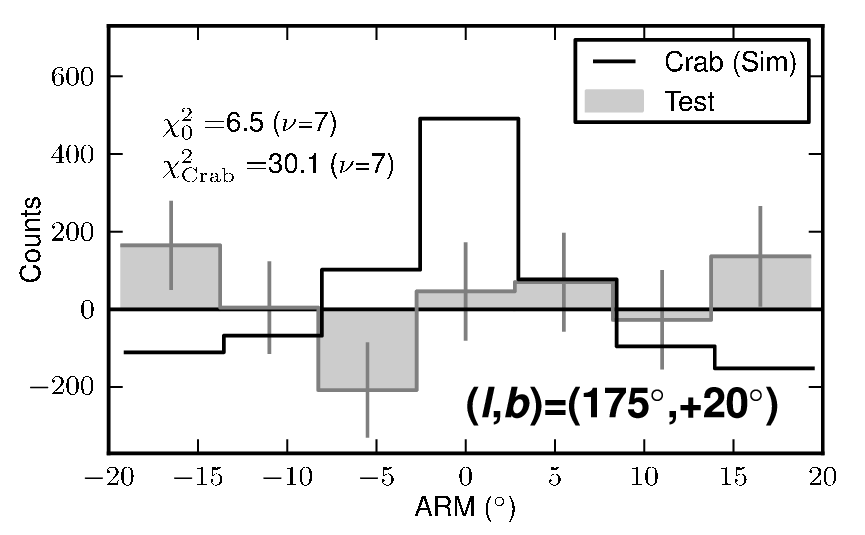}
	\end{center}
	\caption{Measured ARM excess for three different test points,
	with the expected excess for the actual Crab position shown\@. The spectrum
	is measured by subtracting the average spectrum from eight points
	20$^{\circ}$ away from each test point\@. The ARM excess at the three
	test points is consistent with zero\@.\label{fig:arm_flight_testpoints}}
\end{figure}

To test the ARM analysis method, three other test points were chosen that had similar exposure during the flight: (160$^{\circ}$, -10$^{\circ}$), (170$^{\circ}$, +10$^{\circ}$) and (175$^{\circ}$, +20$^{\circ}$)\@. Each of these points is consistent with the null hypothesis (the maximum value of $n_0$ is $1.6\sigma$)\@ and inconsistent with the Crab simulation (the minimum value of $n_{\rm{Crab}}$ is $2.2\sigma$)\@. The results are shown in Fig.~\ref{fig:arm_flight_testpoints}\@.

\section{Conclusions}

\label{sec:conclusions}

The Crab Nebula was detected by NCT at a significance of at least $4\sigma$ during 29.3~ks of observation time in the energy range 300--1500~keV\@.  Compton events with 2--7 individual photon interactions were used as long as the interactions were well-separated in the germanium detectors\@. Using image processing, the Crab appears in a Compton image made from the data\@. Examination of the spectrum and ARM histograms from the source are consistent with Monte Carlo simulations of the Crab observation\@.

This result is the first significant detection of a celestial source by a CCT and is an important step in establishing the viability of the compact Compton telescope design for future space-based wide-survey instruments, such as the proposed Advanced Compton Telescope (e.g.,~\cite{Boggs:2006:ACT})\@.  In the case of ACT, the improvement in efficiency and background rejection of the compact Compton telescope design is predicted to increase sensitivity by as many as two orders of magnitude over COMPTEL in a $10^6$~s observation~\citep{Boggs:2006:ACT}\@.  A similar improvement is predicted over INTEGRAL/SPI, a coded mask instrument that is the most sensitive soft gamma-ray telescope flown to date~\citep{Vedrenne:2003:SPI}\@.

Though NCT should be sensitive to the polarization of gamma-ray sources, a polarization analysis of the Crab was not performed here due to the lack of sufficient statistics\@.  Recent gamma-ray observations have measured a high degree of polarization for the Crab Nebula: $\sim 45\%$ in the 200--800~keV (IBIS) and 100--1000~keV (SPI) regimes~\citep{Forot:2008, Dean:2008}\@.  Given this level of polarization, and the estimated instrumental modulation factor for NCT of $45\%$~\citep{Bellm:2009:p444}, the expected polarization signal from the Crab should be $\sim 20\%$ of the total signal\@.  Therefore, with no other changes to the NCT instrument, an increase in observation time by a factor of $\sim 25$ would be needed to detect the presence or absence of polarization at $4\sigma$ significance, and even more to measure the degree of polarization itself\@.  Further collimation of the instrument could reduce this time, at the cost of a reduced field of view\@.

The NCT instrument suffered extensive damage during a launch attempt in Australia on a subsequent balloon campaign\@.  Once the instrument has been rebuilt, the focus of future balloon flights is to perform polarization analysis on point sources and to image extended sources such as the galactic $^{26}$Al emission\@.

\acknowledgments
The NCT project is funded by NASA under Grant \#NNG04WC38G for the NCT-US team and by the National Space Organization (NSPO) in Taiwan under Grants 96-NSPO(B)-SP-FA04-01 and 98-NSPO-145 for the NCT-Taiwan team\@. The team is grateful to Steve McBride, Jane Hoberman, and C.-H.~Lin for their design of the NCT electronics\@. We would also like to thank the anonymous referee, whose comments greatly clarified this paper\@.


\end{document}